\documentclass[%
 reprint,
 amsmath,amssymb,
 aps,nofootinbib,superscriptaddress, preprintnumbers, longbibliography
]{revtex4-2}

\usepackage{tikz}

\usepackage[utf8]{inputenc}
\usepackage{amsmath,amsfonts,amssymb}
\usepackage{graphicx}
\usepackage[toc,page]{appendix}
\usepackage{cleveref}
\usepackage[loose]{units}
\usepackage{siunitx} 
\usepackage{booktabs}
\usepackage{mathtools}
\makeatletter
\gdef\@fpheader{}
\makeatother

\usepackage{graphicx}
\usepackage{dcolumn}
\usepackage{bm}

\begin{document}

\preprint{}

\title{Lattice simulations of Abelian gauge fields \\coupled to axions during Inflation}

\author{Angelo Caravano}
\affiliation{	Universit\"ats-Sternwarte M\"unchen, Fakult\"at f\"ur Physik, Ludwig-Maximilians Universit\"at, Scheinerstr. 1, 81679 M\"unchen, Germany}
\affiliation{Max-Planck-Institut f\"ur Physik (Werner-Heisenberg-Institut),
	F\"ohringer Ring 6, 80805 Munich, Germany}

\author{Eiichiro Komatsu}%
\affiliation{Max Planck Institute for Astrophysics, Karl Schwarzschild Str. 1, Garching, 85741, Germany}
\affiliation{Kavli Institute for the Physics and Mathematics of the Universe (Kavli IPMU, WPI), University of Tokyo, Chiba 277-8582, Japan}

\author{Kaloian D. Lozanov}

\affiliation{
Illinois Center for Advanced Studies of the Universe \& Department of Physics, University of Illinois at Urbana-Champaign, Urbana, IL 61801, USA.
}%

\author{Jochen Weller}
\affiliation{	Universit\"ats-Sternwarte M\"unchen, Fakult\"at f\"ur Physik, Ludwig-Maximilians Universit\"at, Scheinerstr. 1, 81679 M\"unchen, Germany}
\affiliation{Max Planck Institute for Extraterrestrial Physics,
	Giessenbachstr. 1, 85748 Garching, Germany}

\begin{abstract}
We use a lattice simulation to study a model of axion inflation where the inflaton is coupled to a U(1) gauge field through Chern-Simons interaction. These kinds of models have already been studied with a lattice simulation in the context of reheating. In this work, we focus on the deep inflationary phase and discuss the new aspects that need to be considered in order to simulate gauge fields in this regime. Our main result is reproducing with precision the growth of the gauge field on the lattice induced by the rolling of the axion on its potential, thus recovering the results of linear perturbation theory for this model. In order to do so, we study in detail how the spatial discretization, through the choice of the spatial derivatives on the lattice, influences the dynamics of the gauge field. We find that the evolution of the gauge field is highly sensitive to the choice of the spatial discretization scheme. Nevertheless, we are able to identify a discretization scheme for which the growth of the gauge field on the lattice reproduces the one of continuous space with good precision.
\end{abstract}

\maketitle


\section{Introduction}
In its minimal realization, cosmological inflation is driven by a single scalar field whose negative pressure sources the accelerated expansion of the early universe \cite{PhysRevD.23.347,Sato:1980yn,Linde:1981mu,PhysRevLett.48.1220,STAROBINSKY198099}. Scalar field inflation provides a natural mechanism to generate fluctuations in the energy-density of the universe, which are described as quantum fluctuations of the inflaton field \cite{Starobinsky:1979ty,Mukhanov:1981xt,HAWKING1982295,PhysRevLett.49.1110,STAROBINSKY1982175,Abbott:1984fp} and are observed as temperature fluctuations in the CMB radiation \cite{refId0}. Despite the remarkable observational success of the single scalar field theory, significant effort has been put into trying to embed inflation with gravity into a UV complete description of the early universe. For this purpose, more complicated models of inflation have been considered. 

We consider a nonminimal class of models where inflation is driven by an axion field, which is naturally coupled to gauge fields through Chern-Simons interaction $\phi F\tilde F$. The presence of this interaction is typically associated to an abundant gauge field production induced by the rolling inflaton, which then backreacts on the inflationary dynamics and results in interesting phenomenological signatures in the scalar and tensor perturbations in both Abelian \cite{Anber_2006,Anber_2010,Barnaby_2011_Large, Barnaby_2011, Anber_2012} and non-Abelian \cite{Maleknejad_2011,maleknejad2013gaugeflation,Adshead_2012,Adshead_2013,maleknejad2021su2r} gauge field models. In many cases, these effects are non linear and potentially invalidate the use of linear perturbation theory \cite{Ferreira:2015omg,Peloso_2016,Papageorgiou:2018rfx,Maleknejad_2019,Lozanov_2019,Mirzagholi:2019jeb,Papageorgiou:2019ecb}, making it necessary to employ nonlinear tools to understand their rich phenomenology \cite{2016numst,2016dissax,2020warmde,domcke2020resonant,gorbar2021gaugefield}. For this reason, 
fully nonlinear 
lattice simulations might be a crucial tool in the future in order to compute observables from these models during inflation.  

Various lattice codes have been developed in the last decades to study both scalar \cite{Khlebnikov_1996,Prokopec_1997,latticeeasy,Frolov_2008,hlattice,Sainio_2012,Child_2013,Easther_2010} and gauge field \cite{Lozanov_2020,figueroa2021cosmolattice} models during the reheating epoch and the final $e$-folds of inflation. 
Recently, lattice techniques have also been generalized to the deep inflationary regime for scalar field models \cite{caravano2021lattice,cai2021lattice}. 
However, gauge fields have never been simulated in this regime.

In this work, we use a lattice simulation to study a model of axion inflation where the inflaton is coupled to an Abelian U(1) field during the deep inflationary epoch.
In particular, we aim at reproducing with precision the production of gauge field fluctuations induced by the rolling inflaton through the Chern-Simons interaction. Extending the work of \cite{caravano2021lattice} for the single scalar field case, we are going to reproduce with a lattice simulation the well-known results of linear perturbation theory for this model. For this purpose, we only focus on the gauge field and neglect the backreaction on the dynamics of the inflaton.

 In order to achieve this goal, we study in detail how the discretization scheme, through the choice of the spatial derivatives, influences the dynamics of the gauge field on the lattice. We demonstrate that the choice of the discretization schemes has a profound impact on the dynamics of the gauge field. In particular, we show that standard discretization schemes employed in the literature can lead to an unphysical behavior of the gauge field, whose growth is significantly suppressed on the lattice. Nevertheless, we are able to identity a discretization scheme that is able to reproduce with precision the linear results for this model. This is the main result of this work.
As we explain in more details in the Conclusions, reproducing the linear results is a necessary step before going to the nonlinear regime. Hence, this work represents just a first necessary step in simulating the full axion-U(1) system during inflation. 

This work is organized in the following way. In Section \ref{sec:au1} we briefly introduce the model of axion-U(1) inflation that we consider in this work. In Section \ref{sec:lattice} we introduce the lattice simulation and derive the equations that are used to evolve the fields on the lattice. In this section we also summarize the main differences between our lattice approach and lattice QCD techniques. In Section \ref{sec:disc} we discuss about the choice of the spatial discretization scheme and its consequences on the dynamics of the gauge field. In Section \ref{sec:results} we show the results of the lattice simulation and in \ref{sec:conclusions} we summarize conclusions and future perspectives of this work.



\section{Axion-U(1) Inflation}
\label{sec:au1}
We consider a model of axion inflation where the inflaton is coupled to a U(1) field through Chern-Simons interaction. The action of the early universe in this model is the following:
\begin{align}
\begin{split}
S=\int d^4x \sqrt{-g}\Biggl[&\frac{M^2_{Pl}}{2}R-\frac{1}{2}(\partial_\mu\phi)^2
-V(\phi)\\&-\frac{1}{4}F_{\mu\nu}F^{\mu\nu}-\frac{\alpha}{4f}\phi F_{\mu\nu}\tilde{F}^{\mu\nu}\Biggr],
\label{eq:action}
\end{split}
\end{align}
where $\phi$ is the axion and $V(\phi)$ its potential. $A_\mu=(A_0,\vec{A})$ is the gauge field with $F_{\mu\nu}=\partial_\mu A_\nu-\partial_\nu A_\mu$, $\tilde{F}^{\mu\nu}=\epsilon^{\mu\nu\rho\sigma}/2F_{\rho\sigma}$ and the Levi-Civita tensor $\epsilon_{\mu\nu\rho\sigma}$ is defined such that $\epsilon^{0123}=1/\sqrt{-g}$.  From now on everything will be expressed in reduced Planck mass units $M_{Pl}\equiv1$. 

The background metric is assumed to be the flat Friedmann-Lemaitre-Robinson-Walker\footnote{We set the speed of light $c=1$ here and throughout the rest of this work.}:
\begin{equation}
\label{eq:spacetime}
ds^2=a^2(\tau)(-d\tau^2+d\vec{x}^2).
\end{equation}
We split the inflaton as  $\phi(\vec{x},\tau)=\bar\phi(\tau)+\varphi(\vec{x},\tau)$, while we assume no background value for $A_{\mu}$. We assume a simple slow-roll potential for the inflaton, whose background slowly rolls on the potential so that\footnote{Here and throughout this work, we use the dot to indicate derivatives in cosmic time $dt=a\,d\tau$, i.e. $\dot f=\frac{df}{dt}$, while a prime to indicate derivatives in conformal time $ f^\prime=\frac{df}{d\tau}$} $\dot{\bar\phi}={\bar\phi}^{\prime}/a\ll V(\bar\phi)$. In this work we are interested in studying the growth of the gauge field induced by the slow roll of the background inflaton. In order to see this effect, we first quantize the gauge field as follows:
\begin{align}
\begin{split}
\label{eq:quantization}
\vec{A}(\tau,\vec{x})=\sum_{\lambda=\pm}\int\frac{d^3k}{(2\pi)^{3/2}}\Biggl[\vec{\epsilon}_\lambda(\vec{k})  A_{\lambda}(\tau,\vec{k})a_{\vec{k}}\,e^{-i\vec{k}\cdot\vec{x}}\Biggr] \\ +H.c.,\\	[a_{\vec{k}},a^\dagger_{\vec{k}^\prime}]=\delta(\vec{k}-\vec{k}^\prime),\quad\quad\quad\quad\quad\quad\quad\quad\quad\quad\quad\quad\,\,\,\,
\end{split}
\end{align}
where $H.c.$ means Hermitian conjugate and $\vec\epsilon_\pm$ are the polarization vectors satisfying:
\begin{align}
\begin{split}
\label{eq:epsilon}
&{\vec\epsilon_\lambda}^{\,*}(\vec{k})\cdot 	\vec\epsilon_{\lambda^\prime}(\vec{k})=\delta_{\lambda,\lambda^\prime},\quad \vec{k}\cdot\vec\epsilon_\pm(\vec k)=0,\\ &\vec{k}\times\vec\epsilon_\pm(\vec k)=\mp i k \vec{\epsilon}_\pm(\vec k).
\end{split}
\end{align}
Using this decomposition, one can show starting from the action \eqref{eq:action} and using \cref{eq:epsilon} that the gauge polarizations $A_\pm$
obey the following equations at linear order in perturbations \cite{Anber_2006, Anber_2010}:

\begin{equation}
\label{eq:cont_modes}
A_{\pm}^{\prime\prime}+\left(k^2\pm k \bar\phi^\prime\frac{\alpha}{f}\right)A_{\pm}=0.
\end{equation}
From this equation we can see that the background velocity of the inflaton causes a tachyonic growth of one of the two polarizations of the gauge field for $k<\bar\phi^\prime\alpha/f$. We assume that $ \bar\phi^\prime>0$, so that the growing polarization is $A_-$. Assuming a de Sitter background $\tau=-1/(aH)$ with $H=\dot{a}/a=\text{constant}$, we can rewrite the term in the bracket as $k\bar\phi^\prime\alpha/f=-2\xi/\tau$, where $\xi =\alpha \dot{\bar\phi}/(2fH)$. In this case, assuming Bunch-Davies initial conditions for the gauge field inside the horizon:
\begin{equation}
\label{eq:BD}
A_{\pm}(\vec{k},\tau)=\frac{1}{\sqrt{2k}}e^{- i \omega_{k}\tau}\quad\quad\quad -k\tau\gg 1,
\end{equation}
we can write an exact solution for the growing mode $A_-$ of \cref{eq:cont_modes} \cite{Anber_2006, Anber_2010}:
\begin{equation}
\label{eq:ex_sol}
A_-(k,\tau)=\frac{1}{\sqrt{2k}}\left[G_0(\xi,-k\tau)+iF_0(\xi,-k\tau)\right],
\end{equation}
where $F_\ell$ and $G_\ell$ are the Coulomb wave functions. From these equations we can see that the gauge field experiences a tachyonic growth for modes $-k\tau<2\xi$. This exponential growth will eventually backreact on the inflaton perturbation and will significantly change the 2 and 3 point function of the curvature perturbation $\zeta$ \cite{Anber_2010,Barnaby_2011_Large, Barnaby_2011, Anber_2012}. This puts a constraint on $\xi$, which must be roughly $\xi\sim O(1)$. This means that the exponential growth of the gauge field occurs close to horizon crossing. As mentioned in the introduction, in this work we are not interested in the backreaction effect. Indeed, we will mainly focus on reproducing with a lattice simulation the growth of the gauge field in this observationally interesting parameter range.

\section{Lattice simulation}
\label{sec:lattice} 
In our lattice simulation, we discretize space-time and think of it as a collection of $N^3$ points over a cubic lattice of comoving size $L$ and spacing $\Delta x=L/N$. We associate field values to these points and evolve them using the nonlinear classical equations of motion.

When considering models like the one of \cref{eq:action}, where an Abelian gauge field is coupled to an uncharged scalar, there are different ways of dealing with the discretization procedure. A first approach is to write a discretized action that enjoys a discretized version of the gauge symmetry through the use of \textit{link variables} \cite{Figueroa_2018,Cuissa_2019,Figueroa_2019,figueroa2021cosmolattice}. A second approach, that we follow in this work, is to discretize the system directly at the level of the equations of motion\footnote{These two approaches are equivalent in the case of the axion-U(1) model considered in this work, as it was shown in \cite{2017nonab}.} \cite{Deskins_2013,Adshead_2015,Adshead_2016,Adshead_2018,Braden_2010,2020axiin,2020axiin2,2021axiin}. Following this approach, we derive the equations of motion in a given gauge and check \textit{a posteriori} that the gauge constraint is satisfied during the evolution. In the temporal gauge $A_0=0$, the Euler-Lagrange equations of motion for the axion-U(1) model of \cref{eq:action} are the following \cite{Anber_2010}:
\begin{align}
\begin{split}
\label{eq:eoms}
&		\phi^{\prime\prime}+2H{\phi}^\prime-\nabla^2\phi+a^2\frac{\partial V}{\partial \phi} =-\frac{\alpha}{af}\epsilon_{ijk}A_i^\prime\partial_jA_k\\
&({\partial_j} {A_j})^\prime-\frac{\alpha}{f}\epsilon_{ijk}\partial_k\phi \partial_iA_j=0\\
&{A}^{\prime\prime}_i-\nabla^2A_i+\partial_i({\partial_j} {A_j})\frac{\alpha}{f}\epsilon_{i jk}A_k^\prime\partial_j\phi -\frac{\alpha}{f}\epsilon_{i jk}\phi^\prime \partial_jA_k=0
\end{split}
\end{align}
where $i,j,k\in\{1,2,3\}$ and repeated indices are summed over. Here $\epsilon_{ijk}$ is the totally antisymmetric tensor such that  $\epsilon_{123}=1$, while $\nabla^2$ denotes the Laplacian operator $\nabla^2\equiv\partial_j\partial_j$.

As stated in the Introduction, we are only interested in reproducing the growth of the gauge field induced by the background velocity of the rolling inflaton. For this reason, we will neglect all the terms in the equations of motion that involve spatial derivatives of the axion. We can then use the second equation of \eqref{eq:eoms} to set ${\partial_j} {A_j}=0$. Moreover, we neglect the right-hand side of the first equation because we are not interested in studying the backreaction of the gauge field on the perturbation of the inflaton. In the end we are left with the following equations of motion on the lattice:
\begin{align}
\label{eq:lateoms1}
&		\partial_0^2\phi+2H\partial_0{\phi}+a^2\frac{\partial V}{\partial \phi} =0\\
\label{eq:lateoms2}
&\partial_0^2{A}_i(\vec{n})-[\nabla^2A_i](\vec{n})-\frac{\alpha}{f}\epsilon_{i jk}\partial_0\phi [\partial_jA_k](\vec{n})=0
\end{align}
where the inflaton is spatially homogeneous and the gauge field is a function of the lattice point $\vec n=\{n_1,n_2,n_3\}$ with $n_i\in\{1,...,N\}$. We implement periodic boundary conditions on the lattice in order to minimize boundary effects. Note that periodic boundary conditions are compatible with the homogeneity and isotropy of the inflationary universe. 

We assume a monodromy potential for the inflaton $V(\phi)=\frac{1}{2}m^2\phi^2$ \cite{McAllister:2014mpa}, but the exact choice of the potential is not relevant for the purpose of this work as long as it provides the slow-roll motion of the axion. Note that, within this approximation where we neglect the scalar field fluctuations, \cref{eq:lateoms1,eq:lateoms2} are the same also in the Lorentz gauge\footnote{With the only exception that $A_0$ becomes dynamical with a wavelike equation of motion ${A}^{\prime\prime}_0-\nabla^2A_0=0$.} $\partial^\mu A_\mu=0$. For this reason, the aspects discussed in this work also generalize to simulating the axion-U(1) model in this gauge.

The scale factor in \cref{eq:lateoms1} is evolved using the second Friedmann equation:
\begin{equation}
\label{eq:2fried}
a^{\prime\prime}=\frac{1}{3}(\bar\rho-3\bar p)a^3,
\end{equation}
where $\bar \rho$ and $\bar p$ are the mean energy density and pressure contained in the lattice.
We neglect metric perturbations because they do not play a role in the production of the gauge field at this stage.

In the next section we are going to discuss the choice of the discretization scheme for the Laplacian $\nabla^2$ and for the spatial derivative $\partial_j$ appearing in \cref{eq:lateoms2}. Once a choice is made, \cref{eq:lateoms1,eq:lateoms2,eq:2fried} form a closed system of coupled second-order differential equations that we solve numerically with a Runge-Kutta second order integrator\footnote{We also tried a second order staggered leapfrog integrator, and the results of this work are the same in this case.}. The Coulomb constraint ${\partial_j} {A_j}=0$, however, is not respected exactly on the lattice and we need to ensure that it is approximately satisfied during the evolution. This correspond to checking that the gauge symmetry is preserved and that there are no spurious degrees of freedom propagating on the lattice. We will further discuss this in Section \ref{sec:results}.

Before proceeding, let us stress what are the main differences between the lattice approach presented in this section and more conventional lattice QCD (LQCD) techniques. First of all, as already mentioned, we do not require the gauge symmetry to be exactly preserved on the lattice. Another very important difference is the classical nature of the simulation. In LQCQ, one solves the system using the full path integral, which incorporates all the quantum effects. In our simulation we use the classical equations of motion to evolve the system, and the quantum nature of the gauge field is only relevant when generating initial conditions on the lattice (see Section \ref{sec:IC} for details). This semiclassical approximation is the working assumption of all lattice simulations in the context of inflation and preheating. This approach is equivalent to LQCD computations in the limit where the path integral is dominated by the saddle point corresponding to the classical trajectory and neglecting loop contributions, which are typically subdominant during inflation \cite{Weinberg:2005vy}.




\section{Spatial discretization scheme}
\label{sec:disc} 
In this section we discuss about the choice of the discretization scheme for the spatial derivatives and about the initial conditions for the simulation. In order to do so, we first provide an analytical description of the gauge field dynamics on the lattice that will help us identify a consistent discretization scheme.



The starting point is the definition of the gauge polarizations on the lattice $A_{\pm}$, that are obtained from the discrete Fourier transform of the gauge field as follows:

\begin{equation}
\label{eq:pol}
\vec{A}(\vec{n})=\sum_{\vec{m}}\sum_{\lambda=\pm}\vec{\epsilon}_\lambda(\vec\kappa_{\vec{m}})A_\lambda(\vec\kappa_{\vec{m}})\text{ }e^{-i\frac{2\pi}{N}\vec{m}\cdot\vec{n}},
\end{equation}
where $\epsilon_\pm$ are the polarization vectors defined in \cref{eq:epsilon} and $\vec m=\{m_1,m_2,m_3\}$ with $m_i\in\{1,...,N\}$.
The mode fields $A_{\pm}$ live on the reciprocal lattice where we associate to each point the following comoving momentum:
\begin{equation}
\label{eq:modes}
\vec{\kappa}_{\vec{m}}=\frac{2\pi}{L}\vec{m}.
\end{equation}
We now proceed showing the effects of the discretization on the evolution of $A_\pm$.

\subsection{Consequences of  the discretization on gauge field dynamics}
\label{sec:consequences}
\begin{figure}
	\centering
	
	\begin{tikzpicture}
	\node (img) {\includegraphics[width=9cm]{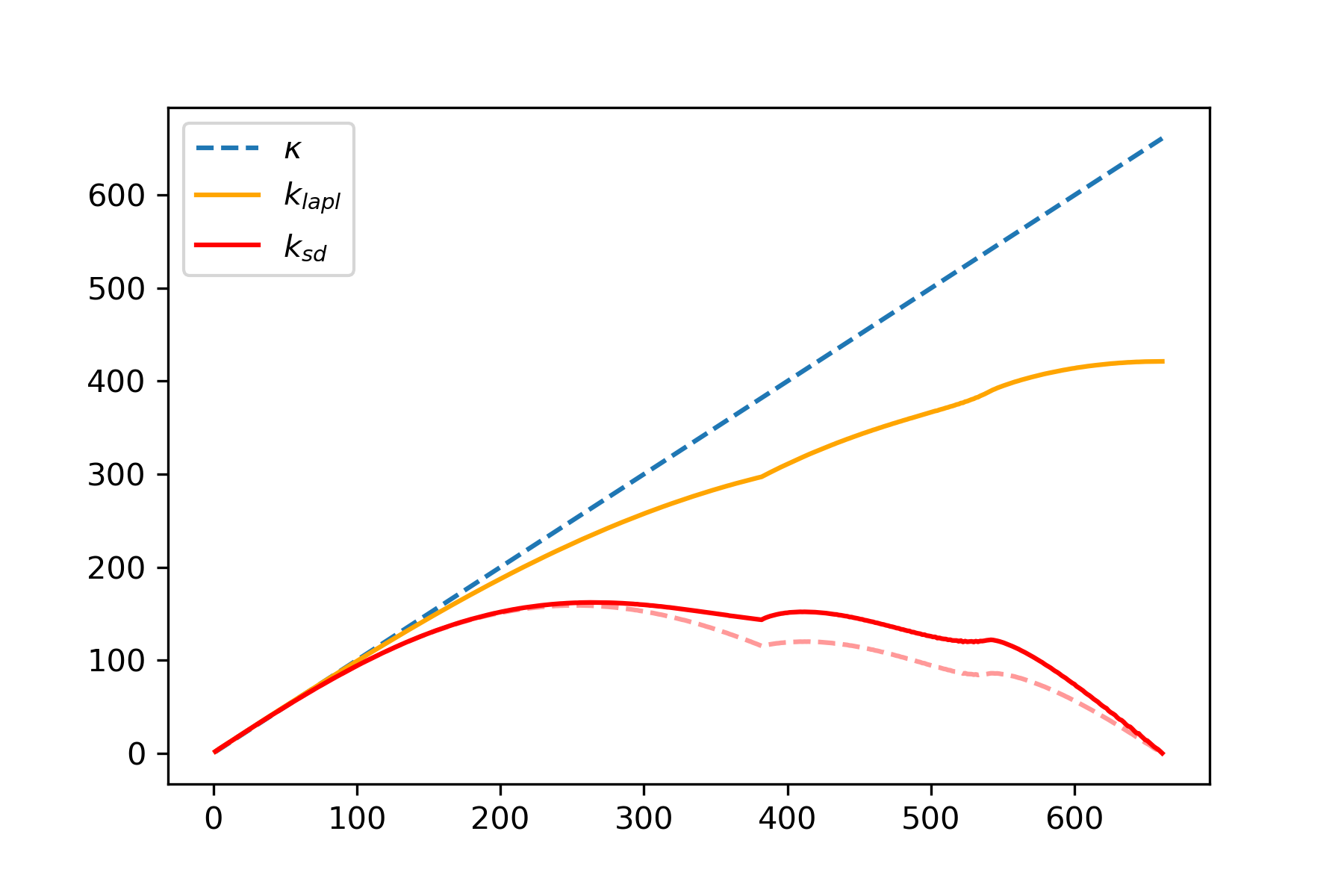}};
	
	\node [text width=0.01cm,align=center] at (0,-3){ $\kappa$};

	\end{tikzpicture}

	\caption{Plot of the different effective momenta emerging from the definitions of the second order centered Laplacian of \cref{eq:lapl_old} (yellow) and the second order centered spatial derivative of \cref{eq:1d} (red). These one dimensional quantities are obtained from \cref{eq:modes,eq:klapl,eq:ksd} through a spherical binning. The red dashed line shows the quantity $\vec{k}_{\rm sd}\cdot \vec{\kappa}/|\kappa|$.}
	\label{fig:dispersion}
\end{figure}

In order to illustrate the issue, let us start with the following standard definitions of lattice Laplacian and one dimensional derivative with second order truncation errors $O(\Delta x^2)$ \cite{press1986numerical}:
\begin{align}
[\nabla^2f](\vec{n})&=\frac{1}{(\Delta x)^2}\sum_{\alpha=\pm1}\biggl(f(\vec{n}+ \alpha\vec{e}_1)\label{eq:lapl_old}\\ &+f(\vec{n}+ \alpha\vec{e}_2)+f(\vec{n}+\alpha \vec{e}_3)-3f(\vec{n})\biggr),\nonumber\\
[\partial_j f](\vec{n})&=\frac{ f(\vec{n}+\vec{e}_j)- f(\vec{n}-\vec{e}_j)}{2 \Delta x} \label{eq:1d}.
\end{align}
where $\vec{e}_1=(1,0,0),\text{ }\vec{e}_2=(0,1,0),\text{ }\vec{e}_3=(0,0,1)$. In Fourier space, these discrete derivative operators result in the following effective momenta:
\begin{align}
\label{eq:klapl}
[\nabla^2 f](\vec{n})\,&\longrightarrow\,{k}^2_{\rm lapl},\quad\vec{k}_{\rm lapl}=\frac{2}{\Delta x}\sin\left({\vec{\kappa}_{\vec{m}}}\frac{\Delta x}{2}\right),\\
\label{eq:ksd}
[\partial_j f](\vec{n})\,&\longrightarrow\,\vec{k}_{\rm sd},\quad\,\,\,\vec{k}_{\rm sd}\text{ }\,=\frac{1}{\Delta x}\sin\left({\vec{\kappa}_{\vec{m}}}\Delta x\right).
\end{align}
With these definitions, we can use the decomposition \eqref{eq:pol} to rewrite \cref{eq:lateoms2} as follows:
\begin{align}
\begin{split}
\label{eq:eomtogh}
\vec{\epsilon}_+A_{+}^{\prime\prime}+&\vec{\epsilon}_-A_{-}^{\prime\prime}+k^2_{\rm lapl}(\vec{\epsilon}_+A_++\vec{\epsilon}_-A_-)\\+&i\frac{\alpha}{f}\phi^{\prime}\vec{k}_{\rm sd}\times(\vec{\epsilon}_+A_++\vec{\epsilon}_-A_-)=0.
\end{split}
\end{align}
In order to simplify \cref{eq:eomtogh}, we notice the following identity\footnote{
	This identity can be derived using \cref{eq:epsilon} and the properties of mixed cross and scalar products.
}:
\begin{equation}
\label{eq:id}
\vec{\epsilon}^{\, *}_\pm(\vec{\kappa}) \,\cdot\,\left(\vec{V}\times\vec{\epsilon}_\ell(\vec{\kappa})\right)=\mp i\left( \vec{V}\,\cdot\,\frac{\vec{\kappa}}{|\vec{\kappa}|}\right)\delta_{\ell,\pm},
\end{equation}
which is valid for any three-dimensional vector $\vec V\in \mathbb{C}^3$. We can use this relation with $\vec{V}=\vec{k}_{\rm sd}$ to obtain two separate equations for the polarization modes of the gauge field $A_{\pm}$. Indeed, multiplying \cref{eq:eomtogh} by $\vec{\epsilon}_\pm(\vec{\kappa})^\ast$ we obtain:
\begin{equation}
\label{eq:latt_modes}
A^{\prime\prime}_\pm+\left(k^2_{\rm lapl}\pm\frac{\alpha}{f}\phi^{\prime}\vec{k}_{\rm sd}\,\cdot\,\frac{\vec{\kappa}}{|\vec{\kappa}|}\right)A_\pm=0.
\end{equation}
This equation is different from its continuous version of \cref{eq:cont_modes}. Indeed, we can see that the main problem with simulating the growth of the gauge field on the lattice are the different effective momenta emerging from the definition of the Laplacian and of the one dimensional derivative. In order to illustrate this effect, let us take a lattice of $N=512$ and $L=4$. In \cref{fig:dispersion} we show a comparison between $\kappa$, $k_{\rm lapl}$ and $k_{\rm sd}$, which are computed as 1-dimensional quantities through a spherical binning on the lattice of \cref{eq:modes,eq:klapl,eq:ksd}. From this plot we can see that $k_{\rm sd}$ (red line) is strongly suppressed with respect to $k_{\rm lapl}$ (yellow line) for most of the scales, and in particular for the largest modes of the simulation.  Indeed, $k_{\rm sd}$ approaches zero for large $\kappa$. From \cref{eq:latt_modes} we can see that this results in an unphysical behavior of the gauge field, which will not be growing on the smaller scales of the lattice\footnote{The reader should keep in mind that we do not use \cref{eq:latt_modes} to evolve fields on the lattice. Indeed, we use the Euler-Lagrange  equations in real space of Section \ref{sec:lattice}, and in this section we are only studying with analytical tools what is expected to happen on the lattice.}. In Section \ref{sec:results} we will discuss in detail the consequences of this effect showing the results of a lattice simulation with spatial derivatives defined as in \cref{eq:lapl_old,eq:1d}.
\subsection{Choice of the discretization scheme}
\label{sec:consistent}

In order to correctly evolve the gauge field on the lattice, we need to find a discretization scheme for the Laplacian and for the one dimensional derivative such that $k_{\rm sd}=k_{\rm lapl}$. Indeed, in this case \cref{eq:latt_modes} will be very similar to its continuous version of \cref{eq:cont_modes}, with the only exception of having $\vec{k}_{\rm sd}\cdot\vec{\kappa}/|\vec{\kappa}|$ inside the bracket instead of $|k_{\rm sd}|$. Later in this section we will discuss about this difference, that turns to be negligible in the evolution of the gauge field. In order to achieve our goal, we define the Laplacian in the following way:
\begin{align}
\begin{split}
\label{eq:lapl}
[\nabla^2 f](\vec{n})=&\frac{1}{(2\Delta x)^2}\sum_{\alpha={\pm2}}\biggl(f(\vec{n}+ \alpha\vec{e}_1)+\\&f(\vec{n}+ \alpha\vec{e}_2)+f(\vec{n}+\alpha \vec{e}_3)-3f(\vec{n})\biggr).
\end{split}
\end{align}
This choice corresponds to defining the Laplacian in a consistent way with respect to the one dimensional spatial derivative. Indeed, once we fix \cref{eq:1d} as one dimensional derivative, this expression for the Laplacian follows by requiring that $\nabla^2=\partial_j \partial_j$ on the lattice, i.e. $[\nabla^2 f](\vec{n})=\left([\partial_j f](\vec{n}+\vec{e}_j)-[\partial_j f](\vec{n}-\vec{e}_j)\right)/(2\Delta x)$. As the Chern-Simons interaction $\phi F\tilde F$ involves spatial derivatives, this will turn out to be a good feature in simulating this model because it allows to consistently compare the Laplacian term $\nabla ^2 A$ and the derivative term $\partial A$ in \cref{eq:lateoms2}.


Using the Laplacian of \cref{eq:lapl} together with the definition of the one dimensional derivative of \cref{eq:1d} will result in the same effective momenta:
\begin{equation}
\label{eq:samemodes}
\vec{k}_{\rm eff}\equiv \vec{k}_{\rm lapl}=\vec{k}_{\rm sd}=\frac{1}{\Delta x}\sin\left({\vec{\kappa}_{\vec{m}}}\Delta x\right).
\end{equation}
Note that, contrarily to \cref{eq:lapl_old}, the Laplacian of \cref{eq:lapl} only employs the next to neighboring points of $\vec{n}$, instead of the neighboring points directly. For this reason, we can interpret this choice as thinking of the $N^3$ cubic lattice as an effective lattice with $N^3_{\rm eff}=(N/2)^3$ points, and where the extra intermediate points are only needed to compute spatial derivatives in a way that is consistent with the problem at hand. As a consequence, only roughly the lower half of the Fourier modes will be physical and we will put a hard cutoff on the lattice in order to exclude the upper part of the spectrum. We choose this cutoff to be the value at which $k_{\rm sd}(\kappa)$ starts decreasing (in \cref{fig:dispersion} this happens roughly around $\kappa\simeq 250$). 

Thanks to this equivalence between $k_{\rm sd}$ and $k_{\rm lapl}$, \cref{eq:latt_modes} will be much closer to its continuous counterpart of \cref{eq:cont_modes}. As mentioned, the only difference is the term inside the brackets where, instead of $|\vec{k}_{\rm sd}|$, we have $\vec{k}_{\rm sd}\cdot \vec \kappa/|\vec \kappa|$. 
We checked by solving numerically the linear equation \eqref{eq:latt_modes} that the presence of this scalar product causes a negligible difference in the evolution of the gauge field. Therefore we neglect this effect and assume $|\vec{k}_{\rm sd}|\simeq \vec{k}_{\rm sd}\,\cdot\, \vec{\kappa}/|\kappa|$.
In \cref{fig:dispersion} we show the difference between these two quantities, which are depicted respectively as a red and a red dashed line in the plot. From this plot we can see that the difference between the two is quite small, and it is negligible in the relevant part of the spectrum (below $\kappa=250$). The validity of this approximation will be confirmed by the results of the simulation. However, as we will see in Section \ref{sec:pol}, this approximation can be avoided by using a different definition of the $\epsilon_\pm$ vectors.

In the end, we can use the results of this section to write a solution for the gauge field growth on the lattice in analogy to the continuous case:
\begin{equation}
\label{eq:ex_sol_lat}
A^{\rm(lat)}_-(\kappa,\tau)\simeq\frac{1}{\sqrt{2k_{\rm eff}}}\left[G_0(\xi,- k_{\rm eff}\tau)+iF_0(\xi,- k_{\rm eff}\tau)\right]
\end{equation}
This solution is similar to \cref{eq:ex_sol} but with $k_{\rm eff}$ instead of $k$, and this is why we refer to it as the effective momentum. Indeed, the lattice solution is equivalent to the continuous one via the identification ${k}_{\rm eff}\leftrightarrow k$, in a similar way to what is shown in \cite{caravano2021lattice} for the single field case\footnote{In other words, the gauge field on the lattice evolve differently with respect to continuous space, as $k_{\rm eff} \neq  \kappa$. However, continuous and discrete dynamics within this discretization scheme are equivalent if we interpret $k_{\rm eff}$ as the physical momentum of the lattice simulation. As we will see in Section \ref{sec:results}, this allows to compare lattice spectra to the continuous ones at all scales.}.

As we will see, this identification turns out to be very useful when comparing the results of the linear theory with the lattice ones.

The strategy that we adopted in this section to achieve the same effective momenta is not necessarily unique. Another way, for example, would be to keep the same definition of Laplacian operator and use the following $O(\Delta x^4)$ stencil for the one dimensional derivative:
\begin{align}
\begin{split}
\label{eq:1d_b}
[\partial_j f]^{(4)}(\vec{n})=\frac{1}{12 \Delta x} \Biggl[-\frac{1}{6}f(\vec{n}+\vec{e}_j)+8f(\vec{n}+\vec{e}_j)\\- 8f(\vec{n}-\vec{e}_j)+ \frac{1}{6}f(\vec{n}-\vec{e}_j)\Biggr].
\end{split}
\end{align}
This leads to the following effective momentum:
\begin{equation}
\vec{k}^{(4)}_{\rm sd}=\frac{1}{\Delta x}\left[\frac{4}{3}\sin\left({\vec{\kappa}_{\vec{m}}}\Delta x\right)-\frac{1}{6}\sin\left(2{\vec{\kappa}_{\vec{m}}}\Delta x\right)\right].
\end{equation}
With this choice we still have $\vec k_{\rm lapl}\neq \vec k^{(4)}_{\rm sd}$, but this time we can find a larger range of modes for which $\vec k_{\rm lapl}\simeq \vec k^{(4)}_{\rm sd}$. This range constitutes roughly the lower half of the spectrum, in a similar way to the strategy above. However, this is achieved only approximately, and the $O(\Delta x^4)$ derivative of \cref{eq:1d_b} is computationally more expensive. For these reasons, we prefer to stick to the first strategy. 



\subsection{Initial conditions for the gauge field}
\label{sec:IC}
As stated in the introduction, the growth of the gauge field in the interesting parameter range $\xi\sim O(1)$ occurs close to horizon crossing. For this reason, we initialize the system when the comoving size of the box is smaller than the horizon $L\lesssim aH$. As a consequence, the gauge field is initiated in its Bunch-Davies vacuum of \cref{eq:BD}. The starting point is the definition of the discretized version of \cref{eq:quantization}:
\begin{align}
\begin{split}
\label{eq:disc_quantization}
&\vec{A}(\vec{n})=\sum_{\lambda=\pm}\sum_{\vec{m}}\biggl[\vec{\epsilon}_\lambda(\vec{\kappa}_{\vec{m}})u_{\lambda}(\vec{\kappa}_{\vec{m}})a_{\vec{m}}e^{-i\frac{2\pi}{N}\vec{n}\cdot\vec{m}}+h.c.\biggr], \\ &[a_{\vec{m}},a^\dagger_{\vec{m}^\prime}]
=\frac{1}{L^3}\delta(\vec{m},\vec{m}^\prime).
\end{split}
\end{align}
The main ingredients are the discrete quantum mode functions $u_{\pm}$, that we write in the following way:
\begin{equation}
\label{eq:discmodes}
u_{\pm}(\vec{\kappa}_{\vec{m}})=\frac{L^{3/2}}{\Delta x^3}\frac{1}{\sqrt{2k_{\rm eff}}}e^{- ik_{\rm eff}\tau }.
\end{equation}
In this expression, we used $k_{\rm eff}$ instead of $\kappa$ so that the initial conditions of the simulation are compatible with the lattice solution of \cref{eq:ex_sol_lat}.  The extra normalization factor $L^{3/2}$ is introduced to correct for the finite volume of space \cite{caravano2021lattice,latticeeasy,Frolov_2008}, while the $1/\Delta x^3$ factor takes into account the dimensionality of the lattice in the computation of the discrete Fourier transform \cite{latticeeasy,caravano2021lattice}.
Since the growth of the gauge field occurs only approximately at horizon crossing, a few modes will already be tachyonic at the beginning of the simulation. For this reason, we will initiate $u_-$ taking into account some of the tachyonic growth, i.e.
\begin{align}
\label{eq:inmodes}
\begin{split}
u_{-}(\vec{\kappa}_{\vec{m}})=\frac{L^{3/2}}{\Delta x^3}\frac{1}{\sqrt{2k_{\rm eff}}}&\biggl[G_0(\xi,- k_{\rm eff}\tau)\\&+iF_0(\xi,- k_{\rm eff}\tau)\biggr],
\end{split}
\end{align}
where we made use of the solution of \cref{eq:ex_sol_lat} for the discrete dynamics. This expression reduces to \eqref{eq:discmodes} for $-k\tau\gg2\xi$, which will be true for most of the modes. Once the mode functions $u_\pm$ are specified, the field configuration on the lattice is generated in Fourier space as statistical realization of a random process. In this way, the uncertainty associated to the quantum nature of the gauge field is replaced by a statistical uncertainty over the different random realizations of the lattice simulation. This procedure
consists in initiating every classical mode $A_{\pm}(\kappa_{\vec{m}})$ as a Gaussian random number with standard deviation equal to $u_{\pm}(\kappa_{\vec{m}})$. We use the same procedure as \cite{caravano2021lattice}, to which we refer for further details.
After this, the gauge field $\vec{A}(\vec{n})$ in real space is obtained from \cref{eq:pol} making use of the polarization vectors $\vec\epsilon_\pm(\kappa)$ defined in \cref{eq:epsilon}. 

\section{Results of the simulation}
\label{sec:results}
In this section we show the results of the code and compare them to the results of linear perturbation theory. The system is initiated far from the end of inflation. This is determined by the background values of the inflaton, that we set to ${\phi}_{\rm in}=-14.5$ and ${\phi}^\prime_{\rm in}=0.8152m$. Here $m$ is the mass of the inflaton, that we set to $m=0.51\cdot10^{-5}$. With this choice, the system is initiated $53$ $e$-folds before the end of inflation. 

\begin{figure}
	\centering

	\begin{tikzpicture}
	\node (img) {\includegraphics[width=6cm]{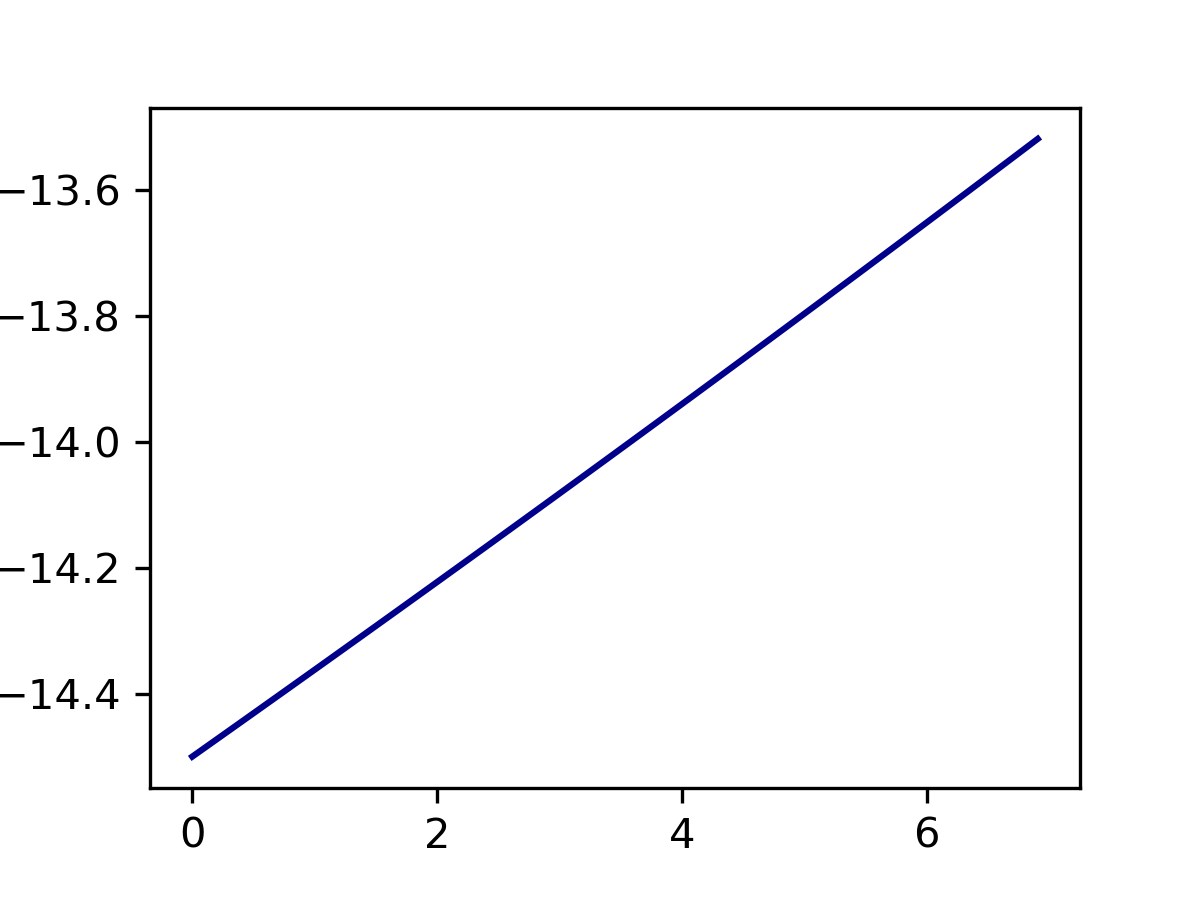}};
	
	\node [rotate=0,text width=0.01cm,align=center] at (-3.5,0){ $\bar{\phi}$};
	\node [text width=0.01cm,align=center] at (0,-2.4){$N_e$};

	\node (img2) at (-0.3,-5)  {\includegraphics[width=6.4cm]{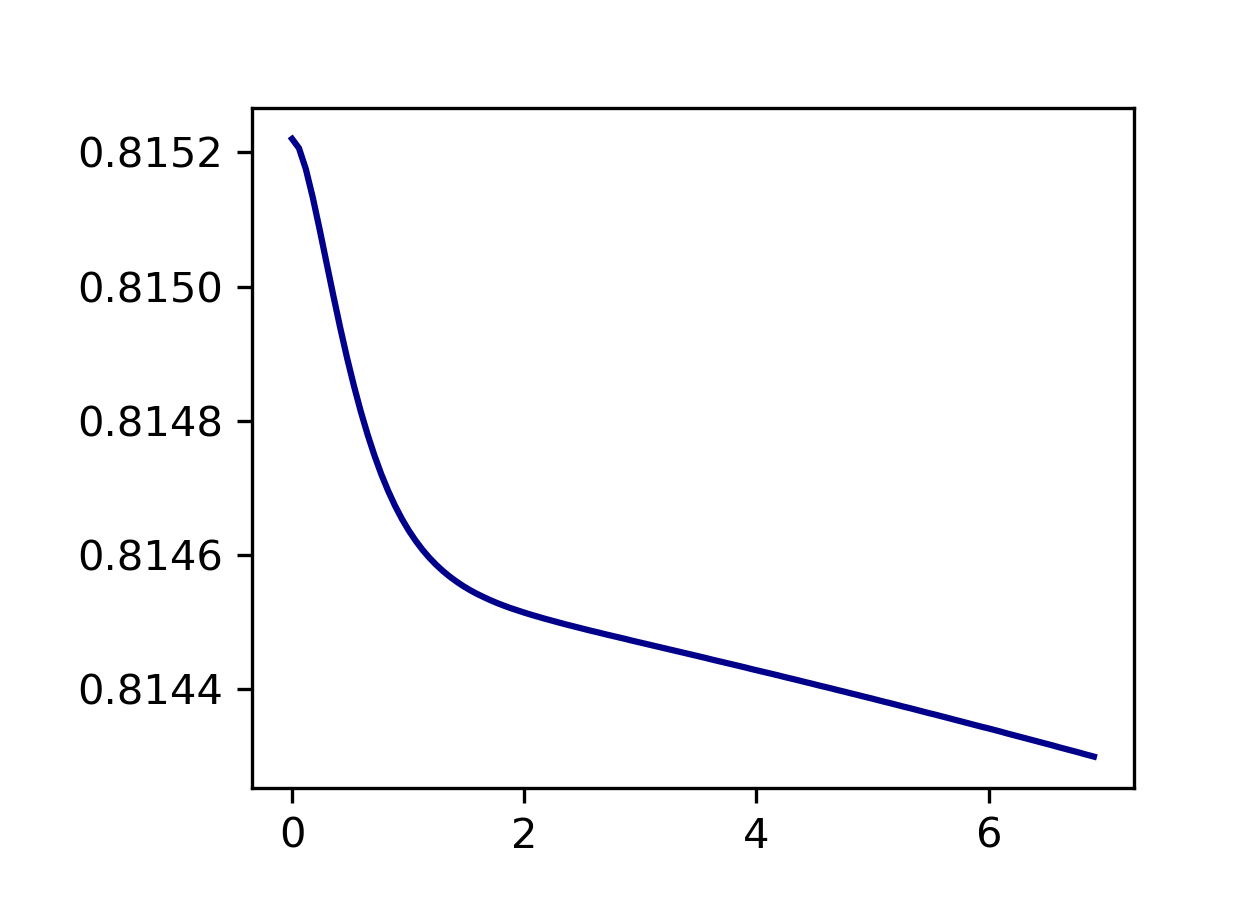}} ;
	
	\node [rotate=0,text width=0.01cm,align=center] at (-4,0-5){ ${\dot{\bar{\phi}}}/{m}$};
	\node [text width=0.01cm,align=center] at (0,-2.4-5) {$N_e$};

	\end{tikzpicture}

	\caption{Plot of background value of the inflaton (top plot) and its velocity (bottom plot) during the simulation.}
	\label{fig:backgroundvalues}
\end{figure}
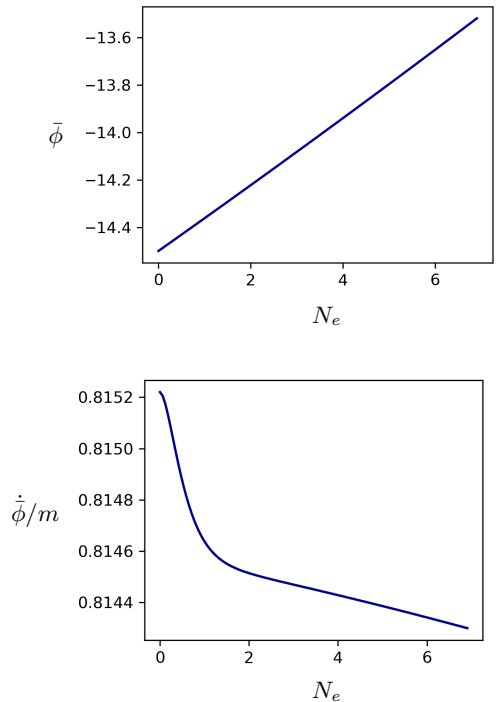

\begin{figure}
	\centering

	\begin{tikzpicture}

	\node (img3) at (0,0) {\includegraphics[width=6cm]{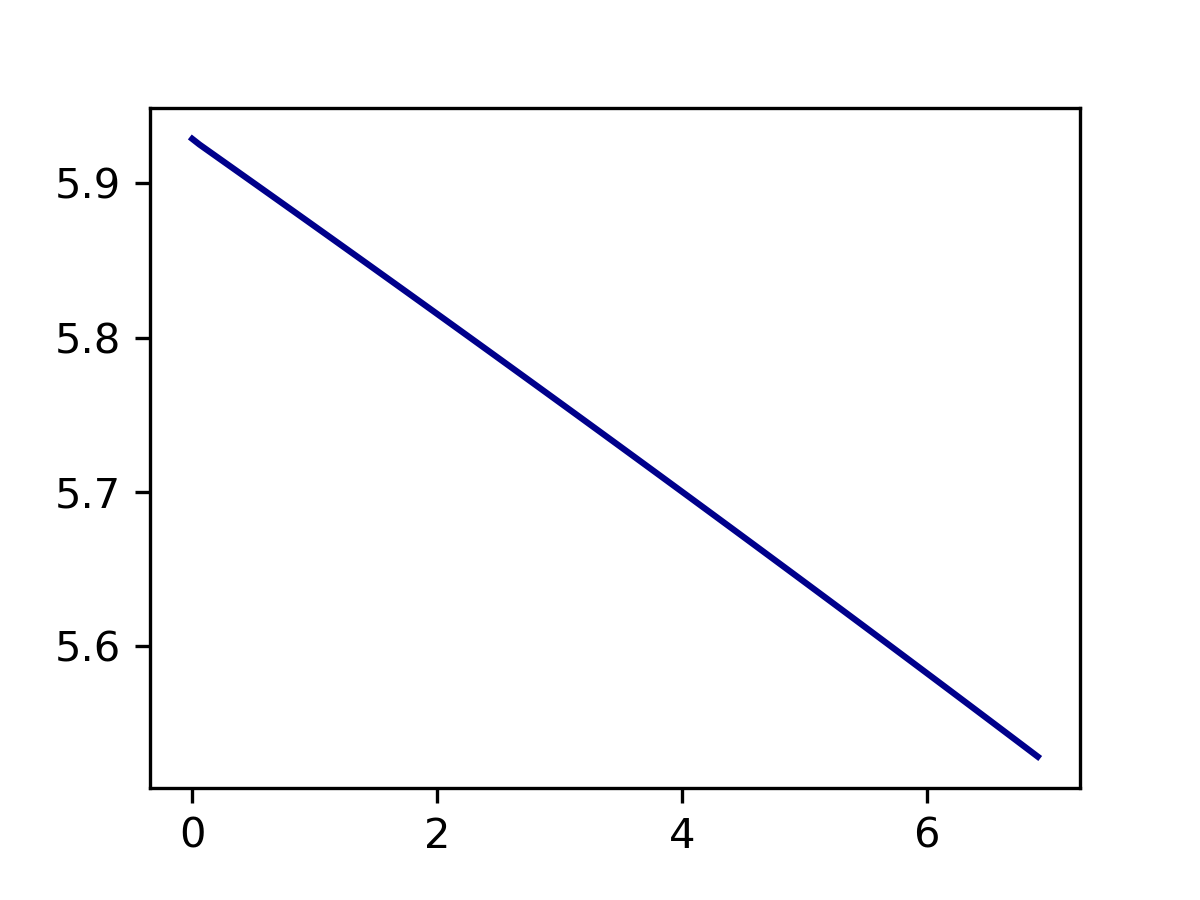}};
	
	\node [rotate=0,text width=0.01cm,align=center] at (-3.8,0){ $H/m$};
	\node [text width=0.01cm,align=center] at (0,-2.4){$N_e$};

	\node (img4) at (-0.3,-5) {\includegraphics[width=6.4cm]{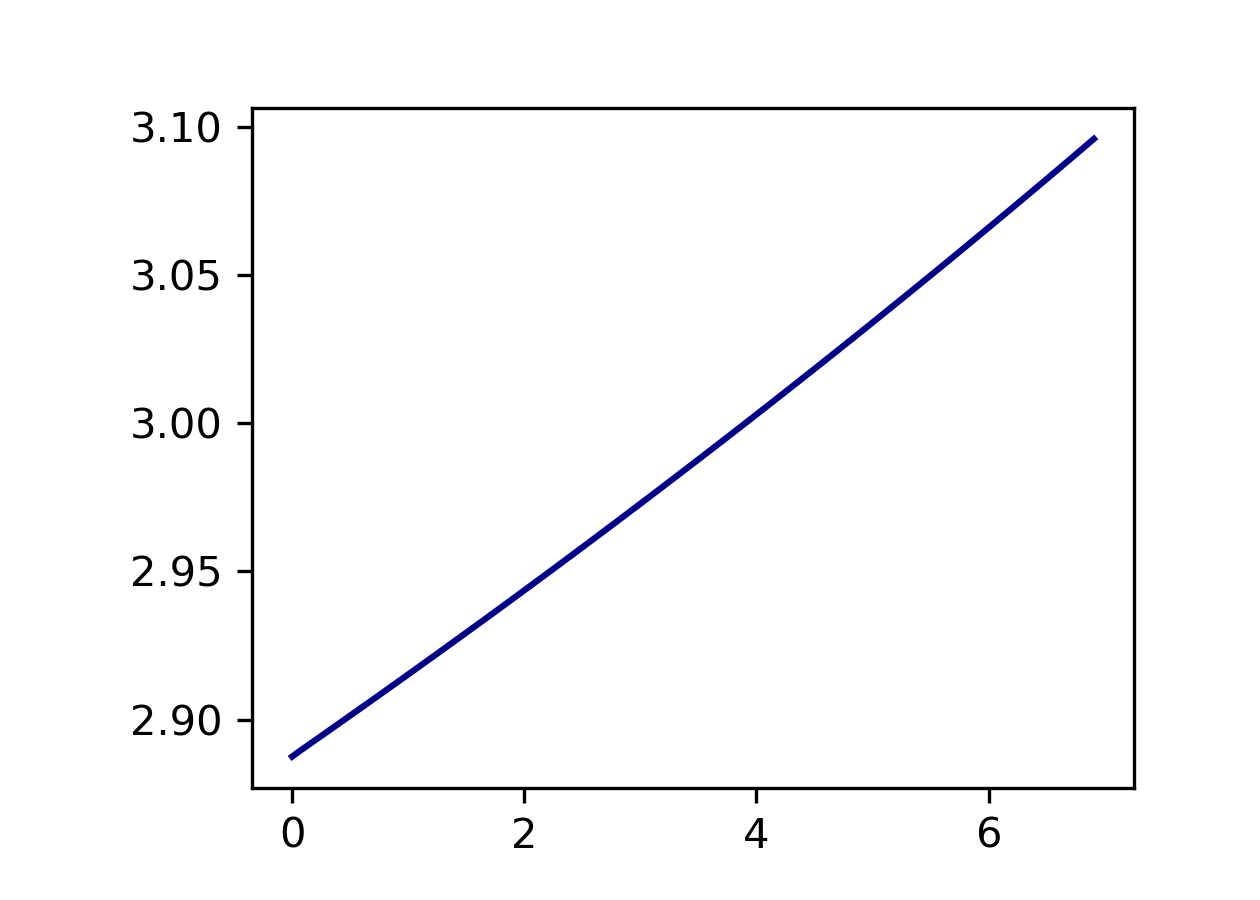}} ;
	
	\node [rotate=0,text width=0.01cm,align=center] at (-3.8,-5){  $\xi$};
	\node [text width=0.01cm,align=center] at (0,-2.4-5){$N_e$};

	\end{tikzpicture}

	\caption{Plot of the Hubble parameter $H$ (top plot) and $\xi$ (bottom plot) during the simulation.}
	\label{fig:backgroundvalues2}
\end{figure}
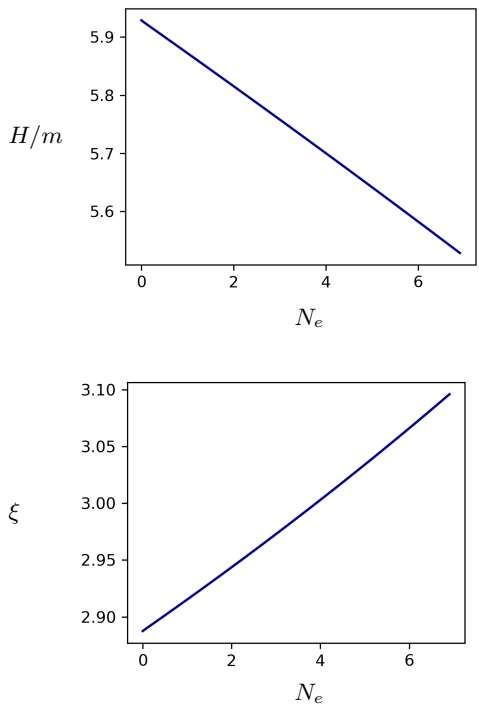

The system is evolved for $N_e\simeq6.9$ $e$-folds from $a=1$ to $a=10^3$ ($N_e=0$ at the beginning of the simulation). In \cref{fig:backgroundvalues} we show the background value of the inflaton and its velocity in cosmic time during the evolution, while the Hubble parameter $H$ is shown in \cref{fig:backgroundvalues2}. 

We take a lattice of $N^3=512^3$ number of points and comoving length $L=4/m$. With these values, the lattice modes $\kappa$ will range from $\kappa_{\rm min}\simeq 0.3 H_i$ to $\kappa_{\rm max}\simeq 118 H_i$, where $H_i$ is the Hubble parameter at the beginning of the simulation.

As commonly done in lattice simulations in the context of inflation \cite{latticeeasy}, we run the simulation with other values of $N$ and $L$ to ensure that our physical outputs (such as the energy density and the evolution of the background quantities) are not influenced by the finite resolution. In this section, however, we only show results with the parameters given above. Note that this is different to what is done in LQCD, where one typically extracts physical quantities by extrapolating lattice outputs in the continuum limit $\Delta x\rightarrow 0 $.

\subsection{Power spectrum}
The main observable that we want to reproduce is the power spectrum of the growing mode of the gauge field. We show results for $\alpha/f=42$, which sets $\xi\simeq 2.9$ at the initial time. Note that the results of this section do not depend on the particular value of the axion-gauge coupling, and we tested the simulation in the range $1<\alpha/f<100$ leading to the same results\footnote{Note that $\alpha/f=42$ with an harmonic potential for the inflaton are ruled out by observations \cite{2020axiin,2020axiin2}. However, as we explain in the text, the results of this section do not depend on the particular value of $\alpha/f$ and we choose this value to better illustrate the growth of the gauge field.}.

\begin{figure*}
	\centering
	
	\begin{tikzpicture}
	\node (img) {\includegraphics[width=15cm]{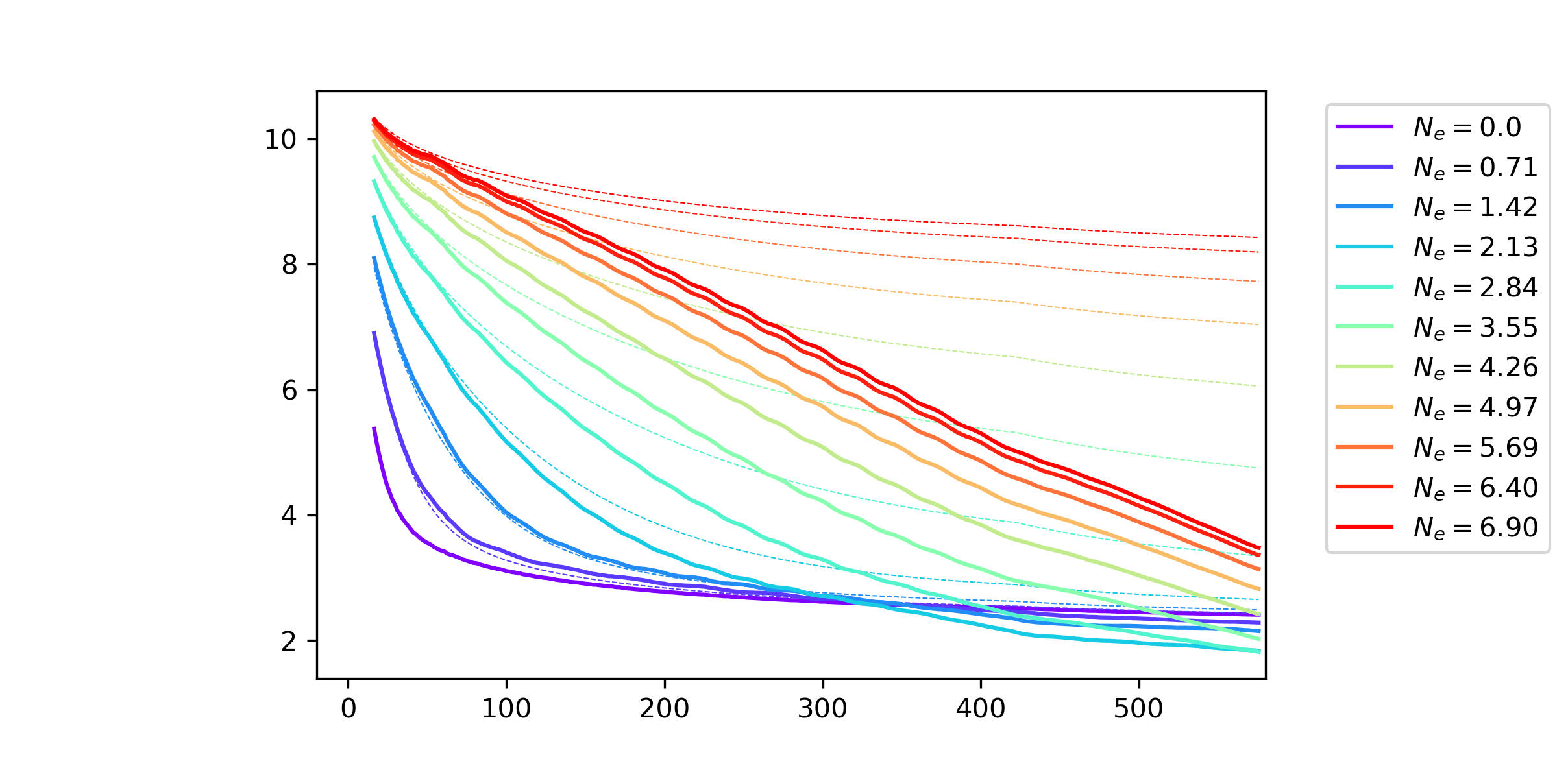}};
	
	\node [text width=0.01cm,align=center] at (0,-3.5){ $\kappa$};
	\node [text width=0.01cm,align=center] at (-7.5,0){ $\log_{10}{|A_-|^2}$};

	\end{tikzpicture}

	\caption{Plot of the power spectrum of the unstable mode of the gauge field, computed from a lattice simulation with a $O(\Delta x^2)$ spatial discretization scheme defined by \cref{eq:lapl_old,eq:1d}. The solid lines represent the power spectrum computed from the simulation at different times, while the dashed lines represent the analytical expectation from the linear theory.}
	\label{fig:gauge_growth_bad}
\end{figure*}

In \cref{fig:backgroundvalues2} we show the plot of $\xi$ during the simulation. In \cref{fig:gauge_growth_bad} we show the power spectrum\footnote{See \cite{caravano2021lattice} for details about how power spectra are computed in our code.} of $A_-$ computed from a simulation with the scheme defined by \cref{eq:lapl_old,eq:1d}, for which $k_{\rm lalp}\neq k_{\rm sd}$. In this plot, the solid lines are the results from the simulation, while the dashed lines represent the expected theoretical power spectrum computed from the linear theory. The theoretical spectra are obtained by solving numerically the linear \cref{eq:cont_modes}, in order to take into account slow-roll corrections to \cref{eq:ex_sol}. From this plot we can see that the simulation is not able to correctly reproduce the growth of the gauge field. Indeed, the gauge field growth is exponentially suppressed with respect to the analytical expectation for most of the modes, affecting both the amplitude and shape of the power spectrum. This expected behavior is a consequence of the sensitivity of the gauge field dynamics to the choice of the spatial discretization scheme, as explained in Section \ref{sec:consequences}. Moreover, the lattice solution shows small oscillations in the form of wiggles at intermediate scales, which are particularly evident at late times (red curves). These oscillations are caused by a misalignment of phase between $A_-$ and its time derivative $A^\prime_-$ at the initial time. This is a consequence of the fact that initial conditions are generated using the continuous solution (i.e. \cref{eq:inmodes}), which is not compatible with the lattice dynamics for this discretization scheme, as described by \cref{eq:latt_modes}.

\begin{figure*}
	\centering
	
	\begin{tikzpicture}
	\node (img) {\includegraphics[width=15cm]{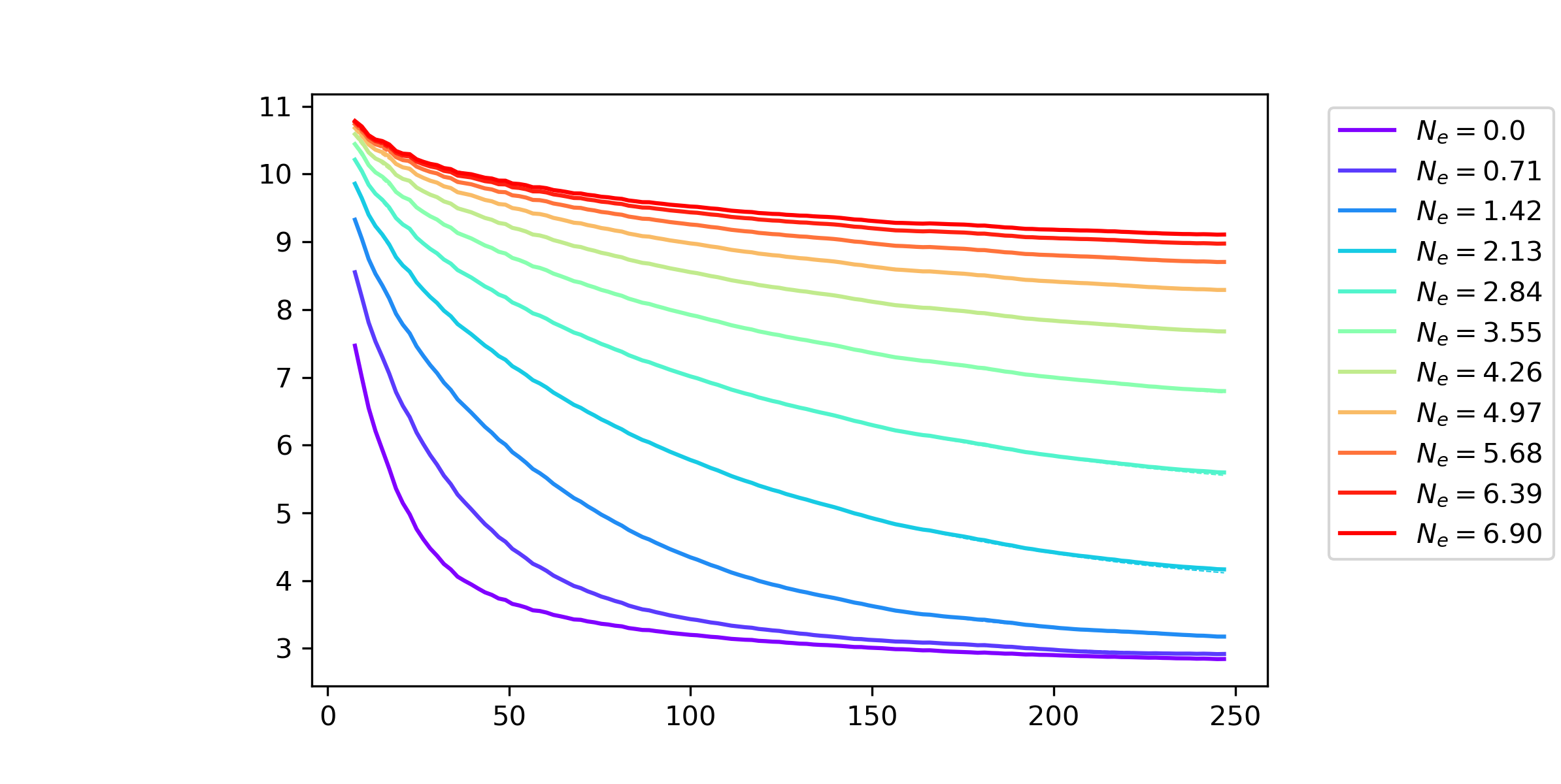}};
	
	\node [text width=0.01cm,align=center] at (0,-3.5){ $\kappa$};
	
	\node [text width=0.01cm,align=center] at (-7.5,0){ $\log_{10}{|A_-|^2}$};
	\end{tikzpicture}

	\caption{Plot of the power spectrum of $A_-$, computed from a lattice simulation with the improved scheme defined by \cref{eq:lapl,eq:1d}. The solid lines are the lattice power spectra, which almost overlap with the expectation from the linear theory depicted as dashed lines (barely visible in this plot).}
	\label{fig:gauge_growth}
\end{figure*}

In \cref{fig:gauge_growth} we show the result from a simulation with the improved scheme where the Laplacian is defined as in \cref{eq:lapl}, for which $k_{\rm eff}=k_{\rm lapl}=k_{\rm sd}$.  We can see that the simulation is now able to reproduce the growth of the gauge field with much better precision. Indeed, the lattice result basically overlaps with the spectra from the linear theory (dashed lines), making them barely visible in the plot.
Within this scheme, $k_{\rm eff}$ plays the role of the effective momentum of the simulation, and for this reason the analytical result in this plot is computed from the linear theory thinking of $k_{\rm eff}$ as the physical momentum (i.e. $k_{\rm eff}$ plays the role of the $k$ of the continuous theory).
As we already discussed in Section \ref{sec:consistent}, the price that we pay when using this discretization scheme is that the upper part of the spectrum is unphysical. For this reason, we set the value of the UV cutoff in this case to be $\kappa_{\rm max}\simeq250/m$, contrarily to $\kappa_{\rm max}\simeq600/m$ of the first scheme above. 

\subsection{Energy density}
\begin{figure*}
	\centering
	
	\begin{tikzpicture}
	\node (img) {\includegraphics[width=11cm]{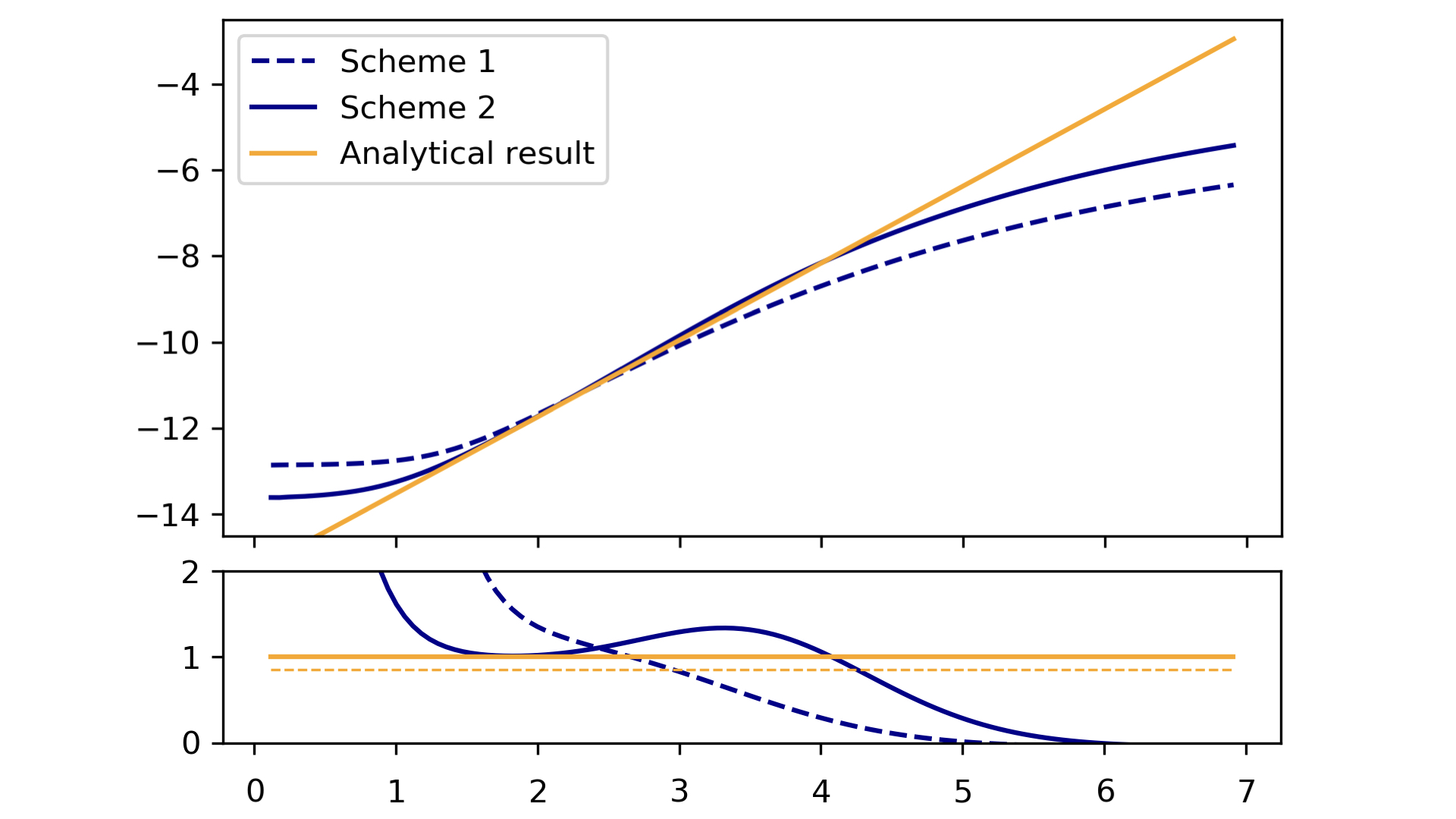}};
	
	\node [rotate=0,text width=0.01cm,align=center] at (-6.6,1.){ $\log_{10}\left(a^4\rho_{\rm GF}\right)$};
	\node [rotate=0,text width=1cm,align=center] at (-5.5,-1.85){ $\rho_{\rm GF}/\rho^{\rm(a)}_{\rm GF}$};
	\node [text width=0.01cm,align=center] at (0,-3.4){$N_e$};

	\end{tikzpicture}

	\caption{Plot of the average value of $\rho_{\rm GF}$ on the lattice. In the top panel we show the value of $\rho_{\rm GF}$, and in the bottom panel we show the same quantity normalized by the analytical result. We call {Scheme 1} the one defined by \cref{eq:lapl_old,eq:1d}, and {Scheme 2} the improved one defined by \cref{eq:lapl,eq:1d}.}
	\label{fig:rhoGF}
\end{figure*}
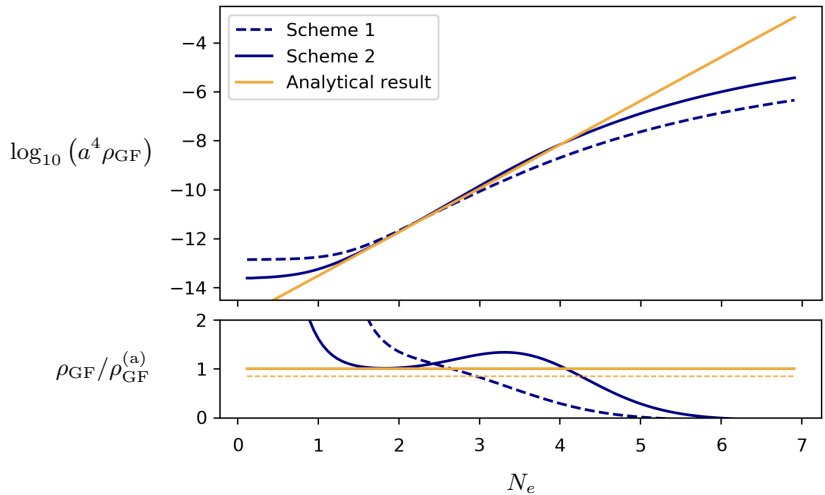

\begin{figure*}
	\centering
	
	\begin{tikzpicture}
	\node (img) at (1,0) {\includegraphics[width=11cm]{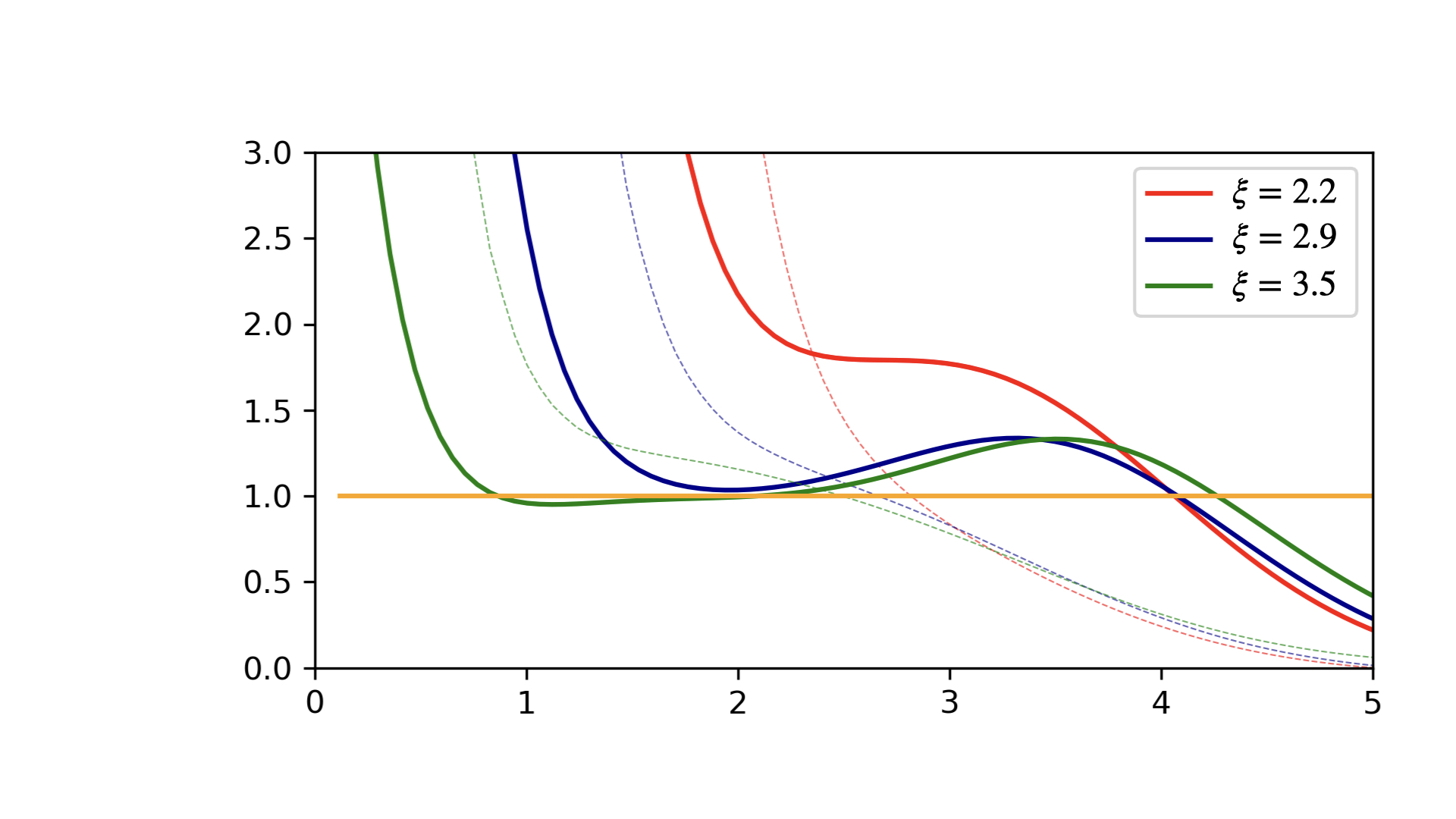}};
	
	
	\node [rotate=0,text width=1cm,align=center] at (-4.3,0){ $\rho_{\rm GF}/\rho^{\rm(a)}_{\rm GF}$};
	\node [text width=0.01cm,align=center] at (1.5,-2.7){$N_e$};

	\end{tikzpicture}

	\caption{Plot of the average value of $\rho_{\rm GF}$ on the lattice divided by the analytical prediction, shown for different values of $\xi$. The $\xi$ in the legend are the values at the beginning of the simulation. The full lines are obtained with the improved scheme of \cref{eq:lapl,eq:1d}, while the dashed ones are obtained with the first scheme of \cref{eq:lapl_old,eq:1d}.}
	\label{fig:rhoGF_xi}
\end{figure*}

In this section we show the evolution of the energy density of the gauge field $\rho_{\rm GF}$ during the simulation. We first show results from the same simulation of the last section with $\xi\simeq2.9$. In the upper panel of \cref{fig:rhoGF} we show the evolution of $a^4\rho_{\rm GF}$ computed as an average energy density over the $N^3$ points of the lattice. We show both the results for the first scheme defined by \cref{eq:lapl_old,eq:1d} (dashed blue line) and for the improved scheme of \cref{eq:lapl,eq:1d} (blue line). To check the accuracy of the lattice result, comparing it to an analytical prediction would be valuable. Unfortunately, an exact analytical result does not exist for a time-dependent $\xi$. However, we can compare the lattice results to the ones obtained in the literature for a constant $\xi$, as it is slowly varying during slow-roll. For a constant $\xi\gg1$, the energy density can be approximated as \cite{Anber_2010}:
\begin{equation}
\label{eq:rhogf}
\rho^{\rm(a)}_{GF}\simeq\frac{6!}{2^{19}\pi^2}\frac{H^4}{\xi^3}e^{2\pi\xi}.
\end{equation}
For a finite $\xi$ this value needs to be corrected to account for a rigorous renormalization procedure \cite{Jimenez:2017cdr,Ballardini:2019rqh,Animali:2022lig}. In the upper panel of  \cref{fig:rhoGF} we show the corrected value of \cite{Ballardini:2019rqh} for $ \rho^{\rm(a)}_{GF}$ (orange line), that we avoid writing explicitly\footnote{	
	More precisely, the $\rho^{\rm(a)}_{GF}(\tau)$ in \cref{fig:rhoGF,fig:rhoGF_xi} shows, for each $\tau$, the $\xi$-constant result of \cite{Ballardini:2019rqh} for the corresponding $\xi(\tau)$.
	
	
	
	
}. 
In the bottom panel of the same figure we show $\rho_{\rm GF}$ from the simulation normalized by the same analytical prediction. The dashed orange line in the bottom panel of \cref{fig:rhoGF} shows the ratio between the expression \eqref{eq:rhogf} and the corrected value.

We can see that the evolution of $\rho_{\rm GF}$ from the simulation can be divided into three phases. During the first phase, from the beginning until $N_e\sim1$, the lattice energy density is much higher than the theoretical expectation due to the classical nature of the simulation. This is because the lattice calculation includes subhorizon UV-divergent contributions that are subtracted in the analytical computation due to the renormalization procedure. These UV-divergent contributions are not subtracted in the lattice computation, as this would lead to an unphysical spatial curvature in the Friedmann equations, as pointed out in \cite{caravano2021lattice}. 
 The initial value of $\rho_{\rm GF}$ is bigger in the case of the first scheme, and this is a consequence of the higher UV-cutoff (see the end of the previous section). After this, there is a second phase until $N_e \sim 4$ in which the lattice result is of the same order of the theoretical value (see more below). This is only true for the second scheme, as the first one significantly underestimates the energy density of the gauge field in real space. During the last $e$-folds of evolution (after $N_e\sim 5$), there is a last phase in which the energy density is much lower than the analytical value, and this is due to the finite size of the lattice. Indeed, at the end of the simulation all the modes become superhorizon and the gauge field production is suppressed due to the finite spatial resolution of the lattice.

In \cref{fig:rhoGF_xi} we show the ratio between the energy density from the simulation and the analytical prediction for different values of the coupling $\alpha/f$. We mainly focus on the results from the improved scheme, but we also show results from the first scheme as dashed lines in the plot.
Since $\xi$ changes during the evolution due to slow-roll, we show the initial value of $\xi$ in the legend of \cref{fig:rhoGF_xi}.
From this plot we can see that for $\xi\gtrsim 3$ 
we can always find an intermediate time range (roughly $1<N_e<4$) in which the result of the simulation is roughly constant and it is close to the analytical expectation of \cite{Ballardini:2019rqh}, up to small corrections due to the time-varying $\xi$. This means that, during this intermediate phase, the finite simulation box contains all the relevant excited modes of the gauge field.
The situation is different for smaller values of $\xi$, and this is because the UV-divergent contributions, subtracted in the analytical computation, become more important in this case. This is a consequence of the classical nature of the lattice simulation. Note that the result of \cite{Ballardini:2019rqh} predicts a negative energy density for $\xi<1.5$. Although the lattice result for the energy density is only approximated with respect to the rigorously normalized result, in particular for small $\xi$, we can see that the second scheme performs significantly better than the first one. Indeed, the second one is never able to achieve a roughly constant $\rho_{\rm GF}$, and the energy density of the gauge field is clearly suppressed in this case.

\subsection{Polarization vectors}
\label{sec:pol}

Let us comment now about the definition of the $\epsilon_{\pm}$ vector used to project $\vec A$ into the polarization $A_{\pm}$. As previously mentioned, $\epsilon_{\pm}(\vec\kappa)$ are defined as in continuous space from \cref{eq:epsilon}.
However, having identified $k_{\rm eff}$ as the physical modes of the lattice theory, it might be better to define the polarization states according to $k_{\rm eff} $ instead of $\kappa$. This corresponds to defining the following lattice polarization vectors $\epsilon_{L,\pm}(\vec \kappa)$:
\begin{align}
\begin{split}
\label{eq:epsilon_lat}
&{\vec\epsilon_{L,\lambda}}^{\,*}(\vec\kappa)\cdot	\vec\epsilon_{L,\lambda^\prime}(\vec\kappa)=\delta_{\lambda,\lambda^\prime},\quad \vec{k}_{\rm eff}(\vec\kappa)\cdot\vec\epsilon_{L,\pm}(\vec \kappa)=0,\\ &\vec{k}_{\rm eff}(\vec\kappa)\times\vec\epsilon_{L,\pm}(\vec \kappa)=\mp i k_{\rm eff}(\vec{\kappa}) \,\vec{\epsilon}_{L,\pm}(\vec \kappa).
\end{split}
\end{align}
Note that, if we do so, in the derivation of \cref{eq:latt_modes} one ends up with $|\vec k_{\rm sd}|$ inside the bracket instead of $ \vec{k}_{\rm sd}\,\cdot\, \vec{\kappa}/|\kappa|$, which makes the approximation made in Section \ref{sec:consistent} unnecessary. In other words, in this case we have an exact equality $=$ instead of $\simeq$ in \cref{eq:ex_sol_lat} (exact in the de Sitter approximation, in the same way of \cref{eq:ex_sol}). However, as we see from the results of this section, this approximation does not spoil the accuracy of the lattice simulation, and for this reason we kept the continuous definition of the $\epsilon_{\rm \pm}$ vectors throughout this work. This, however, might not be true in more complicated cases or when simulating the full axion-U(1) system.

\subsection{Gauge constraint} 
\begin{figure*}
	\centering
	
	\begin{tikzpicture}
	\node (img) {\includegraphics[width=9cm]{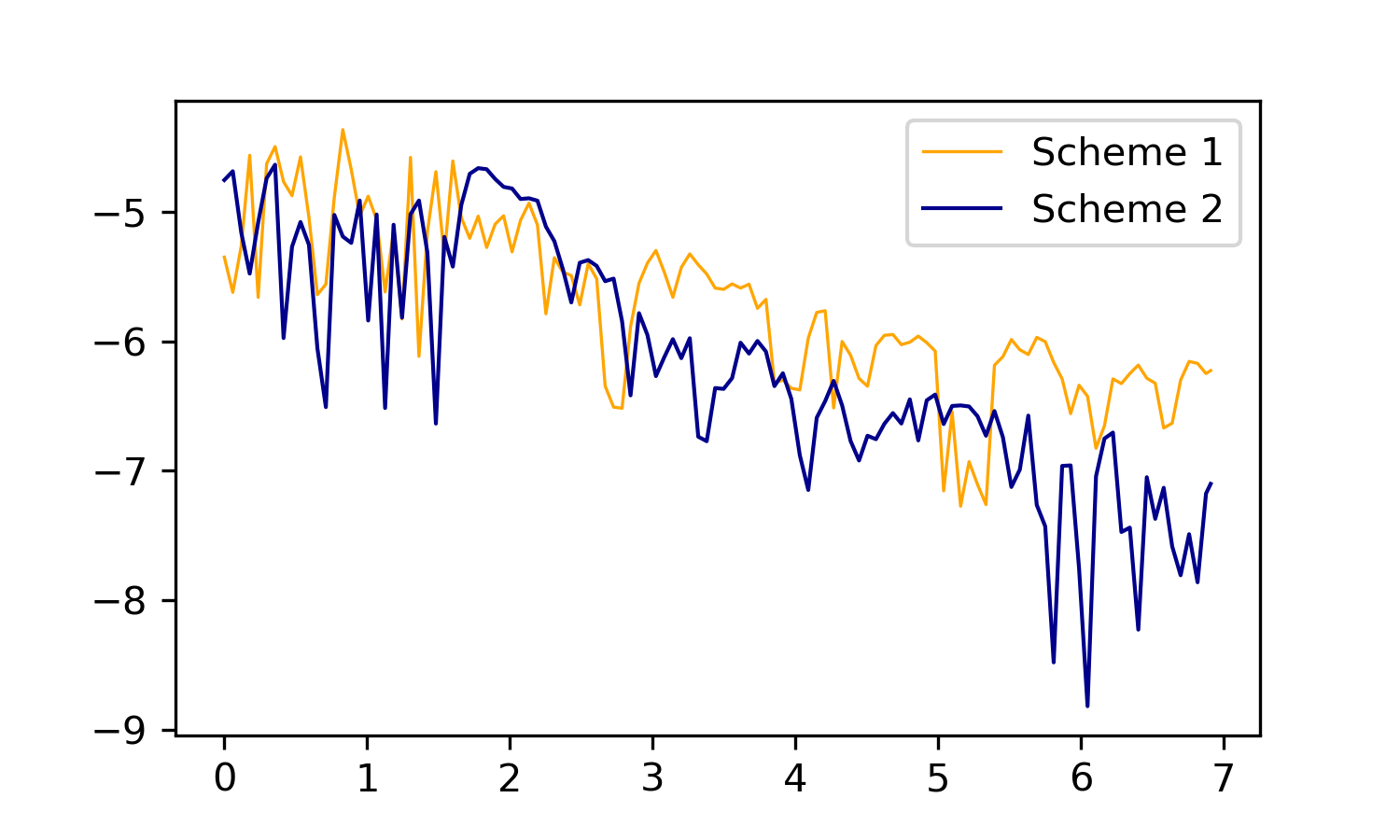}};
	
	\node [text width=0.01cm,align=center] at (0,-2.9){ $N_e$};
	\node [text width=0.01cm,align=center] at (-6.5,0){ $\log_{10}\frac{|\vec \nabla \cdot \vec{A}|}{\sqrt{\sum_i|\partial_i A_i|^2}}$};

	\end{tikzpicture}

	\caption{Plot of the quantity $\vec \nabla \cdot \vec{A}$, normalized by $\sqrt{\sum_i|\partial_i A_i|^2}$, averaged over the lattice as a function of time. We call {Scheme 1} the one defined by \cref{eq:lapl_old,eq:1d}, and {Scheme 2} the improved one defined by \cref{eq:lapl,eq:1d}.
	}
	\label{fig:gauge_constraint}
\end{figure*}

Having reproduced the growth of the gauge field on the lattice, we need to check by hand that the temporal gauge $A_0=0$ is preserved during the evolution. In Section \ref{sec:lattice} we have seen that this corresponds to checking that the Coulomb condition $\vec \nabla \cdot \vec A = 0$ is satisfied. In \cref{fig:gauge_constraint} we plot the average of this quantity over the lattice, normalized by $\sqrt{\sum_i|\partial_i A_i|^2}$ to make it dimensionless. From this plot we can see that the gauge constraint is under control during the simulation. In fact, the violation of the gauge constraint gets smaller during the evolution.

\section{Conclusions and outlook}
\label{sec:conclusions}
In this work we reproduced with a lattice simulation the linear results for an axion-U(1) model of inflation.
In order to do so, we studied with analytical tools how the choice of the spatial derivatives influences the dynamics of the gauge field on the lattice. We found that, in order to reproduce with precision the growth of the gauge field, one needs to find a discretization scheme such that the effective momenta associated to the discrete Laplacian, namely $k_{\rm lapl}$, coincides with the momenta emerging from the one dimentional spatial derivative $k_{\rm sd}$. For standard discretization schemes this is typically not the case, and this can result in a lack of growth of the gauge field, whose power spectrum and energy density are significantly suppressed with respect to the analytical expectation. 
Our main result is finding a new spatial discretization scheme that is able to reproduce with good precision the growth of the gauge field on the lattice, both in terms of power spectrum and in terms of energy density contained in the gauge field.


When dealing with highly nonlinear reheating scenarios, one is mostly interested in qualitative features and a careful comparison between the power spectra coming from the simulation and analytical predictions is typically not relevant. For this reason, the importance of the discretization scheme was never discussed in previous literature. In the case of the deep inflationary regime, however, one is usually interested in computing small deviations from the linear theory. This is why it is important to reproduce with precision the well-known analytical results before performing any computation in the nonlinear regime. 



Having reproduced the results of the linear theory on the lattice, the next natural step would be to study the full axion-U(1) system of \cref{eq:eoms}, allowing for a spatially varying inflaton as well. We leave this for future work. Moreover, the effects described in this work might be relevant also in more complicated models of axion-gauge inflation, where the inflaton is coupled to non-Abelian fields. Indeed, even if the field structure of these models is more complicated, they also manifest a similar gauge field particle production induced by the rolling inflaton \cite{Maleknejad_2019,maleknejad2021su2r}. Therefore, we expect the aspects discussed in this work to be relevant also in simulating these models.

\vspace{0.4cm}
\section*{Acknowledgments}
This work is supported in part by the Excellence Cluster ORIGINS which is funded by the Deutsche Forschungsgemeinschaft (DFG, German Research Foundation) under Germany's Excellence Strategy - EXC-2094 - 390783311 (AC, EK, JW), and JSPS KAKENHI Grant Number JP20H05859 (EK). The Kavli IPMU is supported by World Premier International Research Center Initiative (WPI), MEXT, Japan. The work of KL is supported in part by NASA Astrophysics Theory Grant NNX17AG48G.


\bibliography{U1Sim}

\end{document}